# PHENOMENOLOGICAL DESCRIPTION OF DISORDERING IN FERROELECTRIC MATERIALS CAUSED BY CHARGED DEFECTS.

Anna N. Morozovska[*], Eugeny A. Eliseev[**]

[*]V. Lashkaryov Institute of Semiconductor Physics, NAS of Ukraine,
41, pr. Nauki, 03028 Kiev, Ukraine,

[**]Institute for Problems of Materials Science, National Academy of Science of Ukraine,
Krjijanovskogo 3, 03142 Kiev, Ukraine,
E-mail *morozo@i.com.ua,

## Abstract

We propose the phenomenological description of ferroelectric disordering in ferroelectric-semiconductors caused by charged impurities with the charge density random fluctuations. The material improper conductivity is proportional to the averaged impurity charge density. For high enough conductivity free carriers screen fluctuations in the absence of an external field. When external electric field is applied, inner field fluctuations and dynamical induction fluctuations appear in the inhomogeneously polarized system "charged defect + screening cloud".

We show that the macroscopic state of ferroelectric-semiconductor with random charged defects and sufficiently high improper conductivity can be described by three coupled equations for three parameters averaged over sample volume. Induction determines the ferroelectric ordering in the system, its square fluctuation determines disordering caused by electric field fluctuations appeared around charged fluctuations, and averaged product of fluctuations of induction and charge density reflects the correlations between the free carriers screening cloud and charged defects fluctuations. For the first time we derive the following nonlinear system of coupled equations.

The obtained system of coupled equations adequately describes the peculiarities of polarization switching (footprint and minor hysteresis loops) in such bulk ferroelectric materials as PLZT ceramics and SBN single crystals doped with cerium.

## 1. INTRODUCTION

In most cases stable partial switching of the spontaneous polarization can be achieved in imperfect ferroelectric materials [1]. Sometimes the polarization reversal process is strongly asymmetric even in the bulk samples, and in particular the minor hysteresis loops appear in imperfect ferroelectrics (see e.g. [2], [3], [4], [5]). For the doped ferroelectrics this phenomena is strongly dependent on the type and concentration of dopants [3] and the external field frequency [2], [3]. For the system of sandwich type metal - ferroelectrics - semiconductor [5] this type of switching is defined by the depolarization field and built-in charge layer in the ferroelectrics - semiconductor interface [6]. The similar system was theoretically studied in [7] allowing for the semiconductor properties of ferroelectrics. The model of ref. [7] predicts that only one direction of spontaneous polarization is stable with the ferroelectric layer thickness decrease.

Another interesting feature of the polarization switching in ferroelectrics with non-isovalent dopants are the clamped or pinched hysteresis loops observed in La-doped lead zirconate titanate sol-gel films after the annealing in the hydrogen atmosphere [8]. This type of hysteresis loops are sometimes called footprint loops, their appearance was attributed to the influence of the sample conductivity [9]. Constricted hysteresis loops similar to observed by authors of ref. [8] were found in the La-doped lead zirconate titanate ceramics with composition near the morphotropic boundaries between ferroelectric tetragonal, rhombohedral and antiferroelectric phases [10]. This effect was attributed to the structural changes from cubic matrix with embedded microdomains to orthorhombic macrodomain state twice during one switching cycle of the external field. However we do not know the adequate quantitative model describing the hysteresis loops with constrictions.

We would like to underline that in all aforementioned materials where footprint and minor loops exist either non-isovalent impurities or some unavoidable imperfections manifest itself as charged defects. Moreover, these "dusty" materials would rather be considered as improper semiconductor than ideal dielectrics with random electric fields [11]. These two facts give the basis of our model.

We propose the new phenomenological model, that can give both the simple qualitative explanation and analytical description of partial polarization switching phenomena in the bulk ferroelectric materials with charged defects and high enough improper conductivity. We try to involve the minimum number of hypothesis in our model. Moreover, we have not used the detailed description of the chemical nature, concentration and sizes distribution of randomly situated immovable charged defects, which are the sources of movable charge carriers, inner electric field and induction fluctuations. Also our model admits continuous transformation from the ordered ferroelectric to the disordered material under increasing the charged defects concentration fluctuations.

The main goal of our paper is to demonstrate that macroscopic state of the bulk sample with random charged defects can be described by the system of three coupled equations, which is similar to such well-known nonlinear systems of first order differential equations as the Lorenz one [12]. Such dynamical systems of equations could reveal chaotic regimes, strange attractors as well as strongly non-ergodic behaviour and continuous relaxation time spectrum. Until now we have studied only the stability of the system stationary states by means of the reduced free energy, and simulated quasi-equilibrium ferroelectric hysteresis loops. Certainly, the dynamical dielectric response of this first obtained system requires further investigations.

## 2. THE PROBLEM

We assume that unavoidable charged defects or non-isovalent impurity atoms are embedded into hypothetical "pure" uniaxial ferroelectric. Also we suppose that even in the absence of proper conductivity, impurity atoms provide a rather high improper conductivity in the bulk sample. The concentration of these atoms $\rho_s$ fluctuates due to the great variety of misfit effects (different ionic radiuses, local symmetry breaking, clusterization). These fluctuations ($\delta\rho_s$ are considered in continuous medium framework, i.e. the discrete charge density (point charges in Scheme 1) is approximated by the smooth function $\rho_s = \overline{\rho}_s + \delta\rho_s$. In this approach the short-range fluctuations will be neglected, and the smallest period $d$ in $\delta\rho_s$ spatial spectrum is much greater than the average distance $h$ between real point defect atoms.

Hereinafter we regard embedded impurity centers or defects almost immovable and charged with density $\rho_s(\mathbf{r})$. The sample as a whole is electro-neutral. In this case the movable free charges $n(\mathbf{r},t)$ surround each charged impurity center (see Scheme 1). The characteristic size of these screening clouds is of the same order as the Debye screening radius $R_D$. For the large enough average defect concentration $\overline{\rho}_s$ radius $R_D$ is much smaller than the average distance $h$ between defects. It is obvious, that in the absence of the external electric field $E_0$ the inner field is close to zero outside the screening clouds (see Scheme 1 a,c). But when one applies the external field $E_0$, screening clouds consisting of free charges are deformed, and nano-system "defect center + screening cloud" becomes polarized (Scheme 1 b,d). If the defect charge density fluctuations $\delta\rho_s(\mathbf{r})$ are absent, the short-range electric fields caused by homogeneously distributed induced dipoles are canceled, and no long-range electric field arises in the bulk of the sample (Scheme 1 b). Moreover, the fluctuations $\delta\rho_s(\mathbf{r})$ do not reveal themselves in the absence of the external filed $E_0$ due to the complete screening by movable space-charge fluctuations $\delta n$ (Scheme 1 c). Contrary to this at $E_0 \neq 0$ polarized regions "$\delta\rho_s(\mathbf{r})+\delta n(\mathbf{r},t)$" cause the long-range inner electric field fluctuations $\delta E(\mathbf{r},t)$ (Scheme 1 d). Owing to the equations of state the fluctuations of the inner electric field $\delta E$ cause induction fluctuations $\delta D$. The rigorous conditions that determine the existence of these induction fluctuations will be given below.

Evidently non-homogeneous mechanical stresses appeared near the defects must be taken into account [13]. But the consideration of non-homogeneous mechanical stresses significantly complicates the problem, and we hope that the system behavior would not change qualitatively under the influence of non-homogeneous mechanical stresses. Also it is known (see e.g. [14]) that homogeneous elastic stresses due to the electrostriction coupling with the polar order parameter can be taken into account by the renormalization of the free energy expansion coefficients.

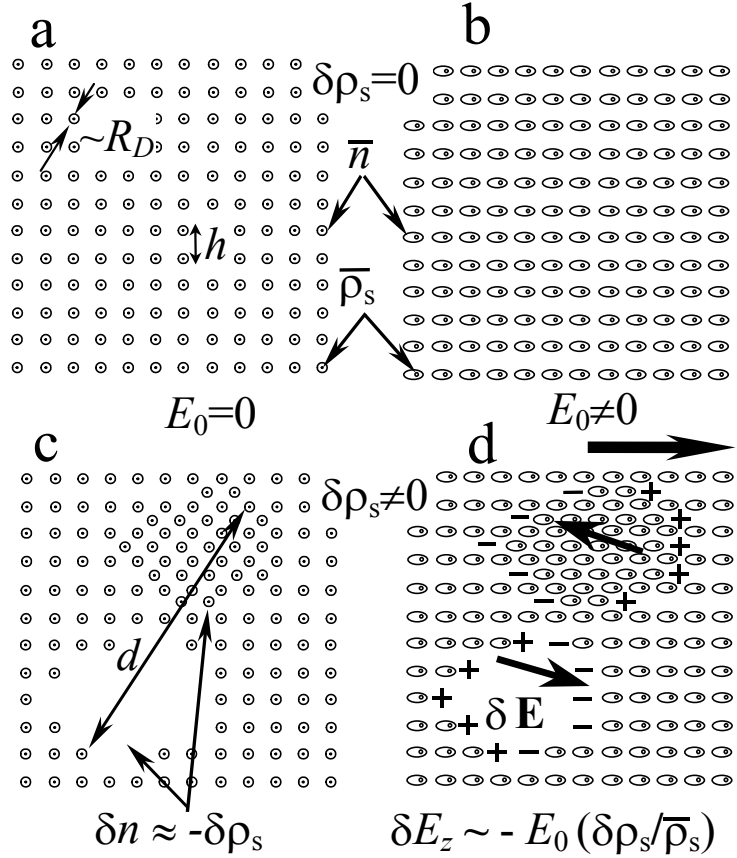

**Scheme 1.** The system of charged defects with the charge density $\rho_s$ (dots) screened by the free charges with density $n$ (circles or ellipses) and screening radius $R_D$. Parts "a", "b" represent charges homogeneous distributions ($\delta\rho_s=0$, $\delta n=0$). Parts "c", "d" represent the distribution with the long-range fluctuations ($\delta\rho_s \neq 0$, $\delta n \neq 0$) and the space period $d$ much greater than the average distance $h$ between defects. The parts "a", "c" and "b", "d" show the system with the zero and nonzero external field $E_0$ respectively.

## 3. GENERAL EQUATIONS

Maxwell's equations for the electric induction $\mathbf{D}$, field $\mathbf{E}$ and equation of continuity have the form:

$$div\,\mathbf{D} = 4\pi\rho, \quad rot\,\mathbf{E} = 0, \quad \frac{\partial\rho}{\partial t} + div\,\mathbf{j}_c = 0 . \tag{1}$$

They have to be supplemented by the equations of state:

$$\mathbf{D}_\perp = \varepsilon_\perp \mathbf{E}_\perp, \quad D_z = D_S + \varepsilon_z(D)E_z, \tag{2}$$

$$\mathbf{j}_c = \sum_m \left( \mu_m \rho_m \mathbf{E} - \kappa_m grad\,\rho_m \right), \quad \rho = \sum_m \rho_m + \rho_s . \tag{3}$$

The sample is regarded as linear dielectric in transverse $x,y$-directions and as nonlinear polar material in longitudinal $z$–direction. Here $\varepsilon_\perp$ is transversal component of dielectric permittivity, $\rho_m$, $\mu_m$ and $\kappa_m$ are the $m$-type movable charge volume density ($m=n,p$), mobility and diffusion coefficient respectively, $\mathbf{j}_c$ is the macroscopic free-carriers current, $\rho_s(\mathbf{r})$ is the given charge density of static defects.

The spatial-temporal distribution of the induction z-component $D_z \equiv D$ can be obtained from the Landau-Ginsburg-Khalatnikov equation:

$$\Gamma \frac{\partial D}{\partial t} + \alpha D + \beta D^3 - \gamma \frac{\partial^2 D}{\partial \mathbf{r}^2} = E_z . \qquad (4)$$

Here $\Gamma > 0$ is the kinetic coefficient, $\alpha(T) = \alpha_T(T - T^*)$, $T$ is the absolute temperature, $T^*$ is the Curie temperature of the hypothetical pure (free of defects) sample, $\beta > 0$, $\gamma > 0$. Equations (1), (2) can be rewritten as:

$$\frac{\partial}{\partial z} D = 4\pi\rho - \varepsilon_\perp \, div\mathbf{E}_\perp ,$$

$$div\left[ \sum_{m=e,h}(\mu_m\rho_m\mathbf{E} - \kappa_m grad\,\rho_m) + \frac{1}{4\pi}\frac{\partial}{\partial t}\left(D\mathbf{e}_z + \varepsilon_\perp\mathbf{E}_\perp\right)\right] = 0. \qquad (5)$$

Here $\mathbf{e}_z$ s the unit vector directed along z-axis.

Hereafter we suppose that homogeneous external field $E_0(t)$ is applied along polar $z$ - axis. The sample occupies the region $-\ell < z < \ell$, i.e. it is infinite in the transverse directions. Let us consider that the electrodes potential difference $\varphi = 2\ell E_0(t)$ at $z = \pm\ell$ which is independent on transverse coordinates. So the inner field satisfies the conditions:

$$\frac{1}{2\ell}\int_{-\ell}^{\ell} E_z(\mathbf{r},t)dz = E_0(t), \quad \int \mathbf{E}_\perp(\mathbf{r},t)\,d\mathbf{r}_\perp = 0. \qquad (6)$$

Boundary conditions depend on the mechanism of the spontaneous induction screening, associates with the formation of oppositely charged space-charge layers with thickness $\ell_c$ [15]. We can assume that the induction distribution is symmetrical for a rather thick sample ($\ell >> \ell_c$) with equivalent boundaries $z = \pm\ell$, i.e.:

$$\frac{1}{S}\int_S D(\mathbf{r}_\perp, z=\ell, t)d\mathbf{r}_\perp \approx \frac{1}{S}\int_S D(\mathbf{r}_\perp, z=-\ell, t)d\mathbf{r}_\perp . \qquad (7)$$

$S$ is the sample cross-section. Also we introduce the averaging over sample volume:

$$f(\mathbf{r},t) = \overline{f(t)} + \delta f(\mathbf{r},t), \quad \overline{f(t)} = \frac{1}{V}\int_V f(\mathbf{r},t)d\mathbf{r} . \qquad (8)$$

Hereinafter the dash designates the averaging over sample volume $V$, $f = \{\rho,\,\rho_m,\,\rho_s, E, D, j, ...\}$, $\overline{\delta f(\mathbf{r},t)} = 0$. All the functions $\delta f(\mathbf{r},t)$ consist of the regular part caused by spontaneous induction screening [15] and the random one caused by fluctuations. Since the contribution from the screening region $|z| > \ell - \ell_c$ to the integrals $\int_V \delta f^n(\mathbf{r},t)d\mathbf{r}$ is negligibly small for the rather thick sample $\ell >> \ell_c$, and $\delta f$ is the fast oscillating function in the remainder of the sample $|z| \le \ell - \ell_c$, one can conclude that:

$$\overline{\delta f^{2n}(\mathbf{r},t)} \sim \left(\overline{\delta f^2(\mathbf{r},t)}\right)^n, \qquad n = 1; 2... \qquad (9)$$

$$\overline{\delta f^{2n+1}(\mathbf{r},t)} \approx 0, \qquad n = 1; 2... \qquad (10)$$

Also we suppose that the correlation between the different $\delta f$-functions is equal to zero if the total power of the functions is an odd number.

It follows from (5), (4) that:

$$E_z(\mathbf{r},t)=E_0(t)+\delta E_z(\mathbf{r},t),\qquad\qquad \mathbf{E}_\perp(\mathbf{r},t)=\delta\mathbf{E}_\perp(\mathbf{r},t),\qquad\qquad(11)$$

i.e. $\overline{E}$ is the applied uniform field $E_0(t)$ and $\overline{\mathbf{E}}_\perp=0$. Notice that the average values $\overline{E},\overline{D}$ are determined experimentally [15], [16] most of the times. Having substituted (6)-(8) into (5) and averaged, one can obtain the expressions for the average quantities, namely:

$$\overline{\rho}=0\;\Rightarrow\;\sum_{m=e,h}\overline{\rho}_m=-\overline{\rho}_s,\quad \overline{\mathbf{j}}(t)=\overline{\mathbf{j}}_c(t)+\mathbf{e}_z\frac{\partial}{\partial t}\frac{\overline{D}(t)}{4\pi}\quad.\qquad\qquad(12)$$

The absence of the space charge average density $\overline{\rho}$ follows from the sample electro-neutrality and corresponds to the result [15], [17]. Here $\overline{\mathbf{j}}(t)$ is the total macroscopic current. Using (5)-(7) one can obtain:

$$\frac{\partial}{\partial z}\delta D=4\pi\left(\sum_{m=e,h}\delta\rho_m+\delta\rho_s(\mathbf{r})\right)-\varepsilon_\perp div\,\delta\mathbf{E}_\perp,\qquad\qquad(13)$$

$$\sum_{m=e,h}\left(\mu_m\left[\delta\rho_m E_0\mathbf{e}_z+\left(\overline{\rho}_m+\delta\rho_m\right)\delta\mathbf{E}-\overline{\delta\rho_m\delta\mathbf{E}}\right]-\kappa_m grad\,\delta\rho_m\right)+\frac{1}{4\pi}\frac{\partial}{\partial t}\left(\mathbf{e}_z\delta D+\varepsilon_\perp\delta\mathbf{E}_\perp\right)=0.\quad(14)$$

Using the nonlinear equation (4), and formula (8)-(11) it is easy to obtain the system of equations:

$$\Gamma\frac{\partial\overline{D}}{\partial t}+\left(\alpha+3\beta\overline{\delta D^2}\right)\overline{D}+\beta\overline{D}^3=E_0(t),\qquad\qquad(15)$$

$$\Gamma\frac{\partial}{\partial t}\delta D+\left(\alpha+3\beta\overline{D}^2\right)\delta D+3\beta\overline{D}\left(\delta D^2-\overline{\delta D^2}\right)+\beta\delta D^3-\gamma\frac{\partial^2\delta D}{\partial\mathbf{r}^2}=\delta E_z.\qquad\qquad(16)$$

The system of equations (13)-(16) is complete, because the quantities $\delta\rho_m,\delta\mathbf{E}$ can be expressed via the fluctuations of induction $\delta D$ and $\delta\rho_s(\mathbf{r})$ allowing for (13), (14). It determines the spatial-temporal evolution of the induction in the bulk sample and has to be supplemented by the initial distributions of all variables.

System (13)-(16) can be used to study the mechanisms of domain wall pinning by the given distribution of charged defects fluctuations $\delta\rho_s(\mathbf{r})$, domain nucleation during spontaneous induction reversal in the ferroelectric semiconductors with non-isovalent impurities. These problems for the ferroelectrics ideal insulators were considered in details earlier (see e.g. [18], [19]). Hereinafter we consider only the average characteristics of the system.

## 4. COUPLED EQUATIONS.

In order to simplify the nonlinear system (13)-(16) the following hypotheses have been used.

a) The sample is the improper semiconductor with rather high $n$-type conductivity:

$$\sum_{m=e,h}\rho_m\approx-\overline{\rho}_s+\delta n,\quad \overline{n}\approx-\overline{\rho}_s,\quad \overline{\delta n}=0,\quad \mu<0,\quad \overline{\rho}_s>0.\qquad\qquad(17)$$

Hereinafter we neglect the proper conductivity and omit the subscript "$m$".

b) The equations (13)-(14) can be linearized with respect to $\delta n$ in the bulk of the sample, where $\delta E\sim\delta n$ and so:

$$\left|\delta n\delta\mathbf{E}-\overline{\delta n\delta\mathbf{E}}\right|<<\overline{\rho}_s\cdot\left|\delta\mathbf{E}\right|.\qquad\qquad(18)$$

c) The characteristic time of $\delta n,\delta\mathbf{E},\delta D$ and $E_0$ changing is the same order as maxwellian time, which is much smaller than Landau-Khalatnikov relaxation time:

$$\frac{1}{4\pi\mu\overline{\rho}_s}<<\frac{\Gamma}{|\alpha|}.\qquad\qquad(19)$$

d) Also we suppose that the smallest period $d$ of the inhomogeneities distribution (see Scheme 1) is much greater then the Debye screening radius $R_D=\sqrt{-\kappa/4\pi\varepsilon_\perp\mu\overline{\rho}_s}$ and correlation length (the thickness of neutral domain wall) $\ell_c=\sqrt{\gamma/\alpha}$, namely

$$R_D^2/d^2 \ll 1, \qquad \ell_c^2/d^2 \ll 1. \qquad (20)$$

Note, that for the typical defects concentration ~1-10% that provides high enough improper conductivity at room temperature $R_D$~5 nm [15], [17], $d$~50 nm, $\ell_c \sim 1\,nm$ [13], [15], i.e. inequalities (20) is valid.

After neglecting the temporal derivatives of $\delta E$ and $\delta D$ (compare with [17]), linearization over $\delta n$ and elementary transformations, the equations (13)-(14) acquire the form:

$$\delta \mathbf{E} \approx \mathbf{e}_z E_0(t) \frac{\delta n}{\overline{\rho}_s} - \frac{\kappa}{\mu \overline{\rho}_s} grad(\delta n), \qquad (21a)$$

$$\delta n - \varepsilon_\perp \frac{\kappa}{4\pi\mu\overline{\rho}_s} \Delta_\perp \delta n \approx \frac{1}{4\pi}\frac{\partial}{\partial z}\delta D - \delta\rho_s. \qquad (21b)$$

The gradient terms in (21b) can be neglected in accordance with (20), because $R_D^2 = -\kappa/\left(4\pi\varepsilon_\perp\mu\,\overline{\rho}_s\right)$ is transverse Debye screening radius (see Scheme 1). Moreover, if only the concentration of free carriers is high enough to provide the good screening of the charged inhomogeneities $\delta\rho_s$, the gradient of the induction fluctuations $(1/4\pi)\partial\delta D/\partial z \cong \delta\rho_s + \delta n$ is small in the bulk of a sample (see Fig. 1), namely:

$$\left|\frac{1}{4\pi}\frac{\partial}{\partial z}\delta D\right| \ll |\delta\rho_s|. \qquad (22)$$

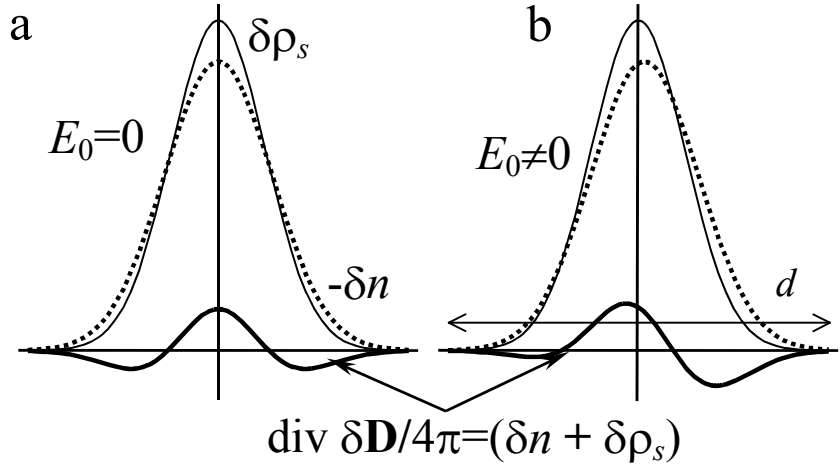

**Figure 1.** The screening of the charged defects $\delta\rho_s$ by free charges $\delta n$.

Thus, the field variation $\delta E_z$ from (21a) can be expressed via $\delta D$ and $\delta\rho_s$ (see Appendix A and [20]):

$$\delta E_z \approx \frac{\kappa}{\mu}\frac{\partial}{\partial z}\left(\frac{\delta\rho_s}{\overline{\rho}_s}\right) - E_0(t)\frac{\delta\rho_s}{\overline{\rho}_s}. \qquad (23)$$

Having substituted solution (23) into (16), we obtain from (15)-(16) the self-consistent system of the nonlinear integral-differential equations for $\overline{D}$ and $\delta D$ Its non-homogeneity is proportional to charge fluctuations $\delta\rho_s$ and external field $E_0$.

The approximate system of first-order differential equations for average induction $\overline{D}$, its square fluctuation $\overline{\delta D^2}$ and correlation $\overline{\delta D\,\delta\rho_s}$ can be derived after some elementary transformations (see Appendix A). Thus we obtain three coupled equations:

$$\Gamma\frac{\partial\overline{D}}{\partial t} + \left(\alpha + 3\beta\overline{\delta D^2}\right)\overline{D} + \beta\overline{D}^3 = E_0(t), \qquad (24a)$$

$$\frac{\Gamma}{2}\frac{\partial}{\partial t}\overline{\delta D^2} + \left(\alpha + 3\beta\overline{D}^2\right)\overline{\delta D^2} + \beta\left(\overline{\delta D^2}\right)^2 = -E_0(t)\frac{\overline{(\delta\rho_s\delta D)}}{\overline{\rho}_s}, \qquad (24b)$$

$$\Gamma\frac{\partial}{\partial t}\overline{\delta D\,\delta\rho_s}+\left(\alpha+3\beta\overline{D}^2+\beta\overline{\delta D^2}\right)\overline{\delta D\,\delta\rho_s}=-E_0(t)\frac{\overline{\delta\rho_s^2}}{\overline{\rho}_s}. \qquad (24c)$$

The system (24) determines the temporal evolution of the bulk sample dielectric response and have to be supplemented by the initial values of $\overline{D}$, $\overline{\delta D^2}$ and $\overline{\delta D\,\delta\rho_s}$ at $t$=0.

Coupled equations (24) have the following physical interpretation (compare with modified approach [20]). The macroscopic state of the bulk sample with charged defects can be described by three parameters: $\overline{D}$, $\overline{\delta D^2}$ and $\overline{\delta D\,\delta\rho_s}$. The long-range order parameter $\overline{D}$ describes the ferroelectric ordering in the system, and the disorder parameter $\overline{\delta D^2}$ describes disordering caused by inner electric fields arising near charged non-homogeneities $\delta\rho_s$. The correlation $\overline{\delta D\,\delta\rho_s}$ determines the correlations between the movable screening cloud $\delta n$ and static charged defects $\delta\rho_s$.

We will show that equations (24) admit the continuous transformation from the perfect ferroelectric ($\delta\rho_s\to 0$ and so $\overline{\delta\rho_s^2}(t)\to 0$) to the local disordering ($\overline{\delta\rho_s^2}\neq 0$ and so $\overline{\delta D^2}(t)\neq 0$) and then to the completely disordered material ($\overline{\delta D^2}(t)>|\alpha|/3\beta$) under the increasing of fluctuations $\delta\rho_s$.

As a resume to this section, we would like to stress that derived non-Hamiltonian system of coupled equations (24) is similar to the other well-known nonlinear systems of first order differential equations (e.g. the Lorenz system). Such dynamical systems posses chaotic regions, strange attractors as well as strongly non-ergodic behaviour and continuous relaxation time spectrum [12]. Any new system of such type demands a separate detailed mathematical study that was not the aim of this paper. Hereafter we discuss only the system behaviour in the vicinity of the equilibrium states far from the possible chaotic regions.

## 5. QUASI-STATIONARY STATES.

Let us consider the stationary solution of (24), which corresponds to the quasi-static external field changing. It is easy to check that the system (24) is not the Hamiltonian one, i.e. it can not be directly obtained by varying of some free energy functional. But in order to study the stability of the stationary points under the external field changing, we try to obtain the reduced free energy.

Let us exclude one of the order parameters from the system (24). Indeed we can easy obtain from (24c) that

$$\overline{\delta D\,\delta\rho_s}=-\frac{E_0\cdot\overline{\delta\rho_s^2}}{\overline{\rho}_s\left(\alpha+3\beta\overline{D}^2(t)+\beta\overline{\delta D^2}\right)}. \qquad (25a)$$

Substituting (25) into (24b) and extracting square root one obtains

$$\sqrt{\overline{\delta D^2}}\cdot\left(\alpha+3\beta\overline{D}^2+\beta\overline{\delta D^2}\right)=\pm E_0\sqrt{\overline{\delta\rho_s^2}/\overline{\rho}_s^2}. \qquad (25b)$$

In the stationary case (24a) acquires the form:

$$\left(\alpha+3\beta\,\overline{\delta D^2}\right)\overline{D}+\beta\overline{D}^3=E_0. \qquad (25c)$$

It is easy to see that system of equations (25b), (25c) can be obtained from variational principle. Really, the equations (25b), (25c) can be integrated over averaged induction fluctuations $\sqrt{\overline{\delta D^2}}$ and induction $\overline{D}$, and so the reduced free energy functional determining the stability of the stationary points has the following form:

$$G\left(\overline{D},\sqrt{\overline{\delta D^2}}\right)=\frac{\alpha}{2}\left(\overline{D}^2+\overline{\delta D^2}\right)+\frac{\beta}{4}\left(\overline{D}^4+\left(\overline{\delta D^2}\right)^2\right)+\frac{3\beta}{2}\overline{D}^2\cdot\overline{\delta D^2}-E_0\left(\overline{D}+\sqrt{\frac{\overline{\delta\rho_s^2\cdot\overline{\delta D^2}}}{\overline{\rho}_s^2}}\right). \qquad (26a)$$

Hereinafter we suppose that the root $\sqrt{\overline{\delta D^2}}$ can be both negative and positive depending on the external conditions. It is easy to verify that the equations of state

$$\frac{\partial}{\partial \overline{D}} G\left(\overline{D}, \sqrt{\overline{\delta D^2}}\right) = 0, \qquad \frac{\partial}{\partial \sqrt{\overline{\delta D^2}}} G\left(\overline{D}, \sqrt{\overline{\delta D^2}}\right) = 0 \qquad (26b)$$

coincide with the equations (25c) and (25b). Let us rewrite energy functional (26a) in dimensionless variables.

$$G_m\left(D_m, \Delta_D\right) = -\frac{1}{2}\left(D_m^2 + \Delta_D^2\right) + \frac{1}{4}\left(D_m^4 + \Delta_D^4\right) + \frac{3}{2} D_m^2 \cdot \Delta_D^2 - E_m\left(D_m + R \cdot \Delta_D\right). \qquad (27)$$

Here $\alpha < 0$, $D_m = \overline{D}/D_S$, $D_S = \sqrt{-\alpha/\beta}$, $E_m = E_0/(-\alpha D_S)$, $\Delta_D = \sqrt{\overline{\delta D^2}}/D_S$, $R = \sqrt{\overline{\delta \rho_s^2}/\overline{\rho_s^2}}$.

The contour plots of negative values of the free energy (27) at different $E_m$ amplitudes and small $R$ value are depicted in Fig.2. The free energy minimums (crosses and stars) prove the existence of the two sorts of stable states: the first one with non-zero averaged induction values $\overline{D} \neq 0$ and $\overline{\delta D^2} \approx 0$ (ordered state), and the second one with its zero value $\overline{D} \approx 0$ and $\overline{\delta D^2} \neq 0$ (disordered state). It is obvious that at zero external electric field all the four states have the same energy, and therefore the system can be in any of these states (see Fig. 2a, c). However when one applies the external field $E_0$ above the definite threshold value, states with induction direction opposite the field one vanish, and system has to switch to the states with induction directed along the external field.

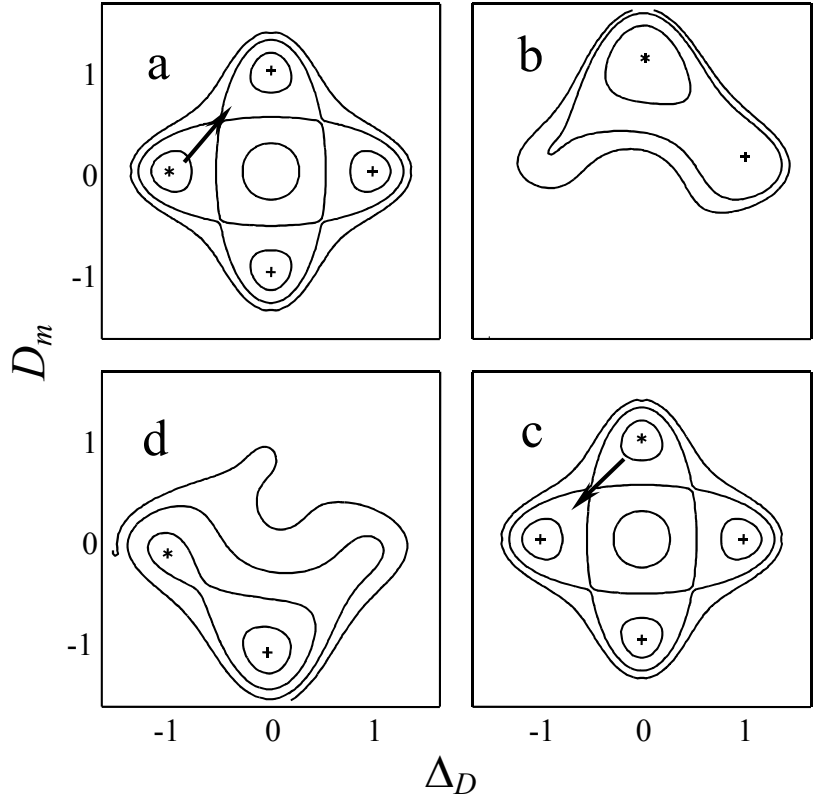

**Figure 2**. The contour plots of negative values of the free energy (27) at $R^2$=0.1 and different values of $E_m$=0, 0.3, 0, -0.3 (parts a, b, c, d respectively). Crosses denote the positions of minimums; star is the position of the current minimum.

In contrast to the pure ferroelectrics, where the switching takes place between two ordered states $\overline{D} = \pm 1$, the considered system from the ordered state $\overline{D} = 1$ (Fig. 2c) switches first to the disordered one $\overline{D} = 0$ (Fig. 2d) and vice versa (Fig. 2a, b). The plots in Fig.2 correspond to the small amplitudes of the external field $E_m$, which can switch the system only between the two adjacent minimums denoted by the star. If the external field $E_m$ is increased, full switching between two ordered states $\overline{D} = \pm 1$ through disordered ones $\overline{D} = 0$ is possible. In such case the field $E_m$ can successively switch the system between four minimums.

The contour plots of the free energy (27) at fixed $E_m$ and different $R$ values are depicted in Fig. 3. Again, free energy minimums speak in favour of existence of the stable ordered and disordered states. The disordered state $\overline{\delta D^2} \neq 0$ is absent in pure ferroelectrics.

The cross-sections of the free energy (27) at $\overline{\delta D^2} = 0$ (completely ordered states) and $\overline{D} = 0$ (completely disordered states) at fixed $E_m$ and different $R$ values are depicted in Fig. 4. As it should be expected, the ordered state is the most energetically preferable than the disordered one at $R^2 \ll 1$ even in the infinitely small external field (see Figs. 3a,b and Fig. 4a,b). One can see that the probabilities of the system existence in ordered and disordered states become very close at $R^2 > 0.5$ (see Fig. 4c). Moreover, at $R \to 1$ ordered and disordered states become energetically indistinguishable (see Fig. 4d).

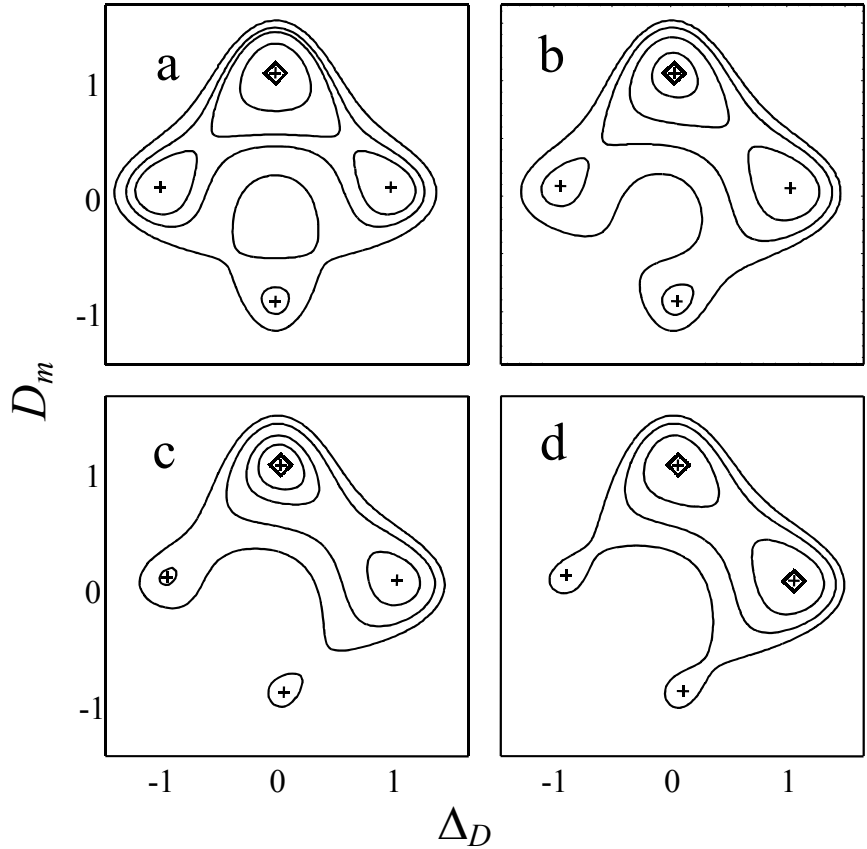

**Figure 3**. The contour plots of negative values of the free energy (27) at $E_m$=0.15 and different values of $R^2$=0, 0.3, 0.5, 1 (parts a, b, c, d respectively). Crosses denote the positions of minimums; marked crosses are the positions of the absolute minimums.

Our calculations proved that at $R > 1$ disordered state with $\overline{D} = 0$, $\overline{\delta D^2} \neq 0$ becomes the most energetically preferable even in the infinitely small external field. Actually this means that the phase transition into disordered state takes place at the critical value of fluctuations $R = 1$, i.e. the ferroelectrics sample with impurities splits into the polar regions (domains or Cross regions [21]) with different induction orientation.

## 6. QUASI-EQUILIBRIUM DIELECTRIC HYSTERESIS.

In this section we demonstrate how the dielectric quasi-equilibrium hysteresis loop $\overline{D}(E_0)$ changes its shape under the presence of charged defects. First of all let us rewrite equation (26) in dimensionless variables:

$$\frac{dD_m}{d\tau} - (1 - 3\Delta_D^2)D_m + D_m^3 = E_m ,$$

$$\frac{1}{2}\frac{d\Delta_D^2}{d\tau} - (1 - 3D_m^2)\Delta_D^2 + \Delta_D^4 = -E_m K_{Dp} , \qquad (28)$$

$$\frac{dK_{Dp}}{d\tau} - (1 - 3D_m^2 - \Delta_D^2)K_{Dp} = -E_m R^2 .$$

Here $\tau = t/(-\alpha\Gamma)$. The dependence of dimensionless induction $D_m$ over the external field $E_m$ is represented in Figures 5-8 for the case of harmonic modulation of the external field $E_m = E_{mA}\sin(w\tau)$. Hereinafter we use the dimensionless frequency $w = -\alpha\Gamma\omega$.

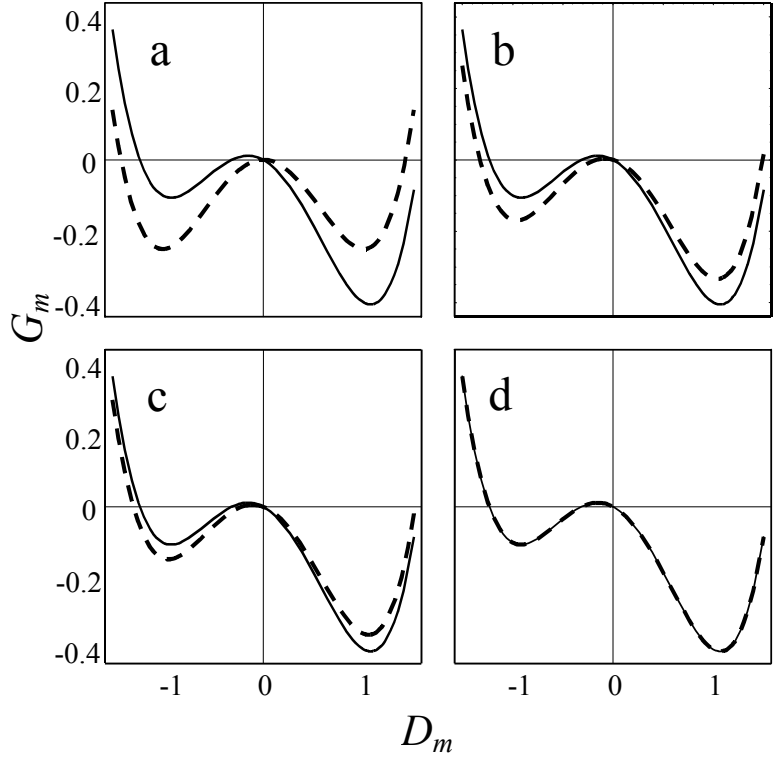

**Figure 4**. The cross-sections of the free energy (27) at $\overline{\delta D^2} = 0$ (dashed curves) and $\overline{D} = 0$ (solid curves) for $E_m = 0.15$ and different values of $R^2 = 0$, 0.3, 0.5, 1 (parts a, b, c, d respectively).

The quasi-equilibrium hysteresis loops were obtained at the low frequency of external field (see Figures 5-6). It is seen from the Figure 5, that a constriction on hysteresis loops appears for the nonzero $R$ values, and the loop area increases with the increase of parameter $R$. The so called "footprint" loops were observed in PLZT ceramics [10]. But for a given external field amplitude there is a critical value of parameter $R$. For the $R$ value above critical one instead of "full" loops we obtained only minor loops (see Fig. 6) or their absence (see Fig. 5d) depending on the ordered ($D_m(\tau=0)=1$) or disordered ($D_m(\tau=0)=0$) initial state of the sample. The similar minor loops were observed in PZT [1] and SBN: Ce single crystals [2].

On the other hand there is a critical value of the external electric field amplitude for the given value of parameter $R$. Also one can see minor loops only for the amplitude of electrical field smaller than critical one (compare Fig. 5 with Fig. 6). Even for the field amplitude below the thermodynamic coercive field ($E_{mc} = 2/3^{3/2} \approx 0.385$) there is the partial switching, when the hysteresis loop is absent in "pure" ferroelectrics (compare dotted and solid curves). It should be noted, that for $R=1$ one can observe only minor loops (see Fig. 6d) for the finite value of the external electric field.

Note that two minor loops in Fig. 6a-d are obtained with several cycles of external field and switching between two loops is caused by the fluctuations. Upper minor loops represent solutions with the positive initial values of the induction, while the lower ones correspond to its negative initial values.

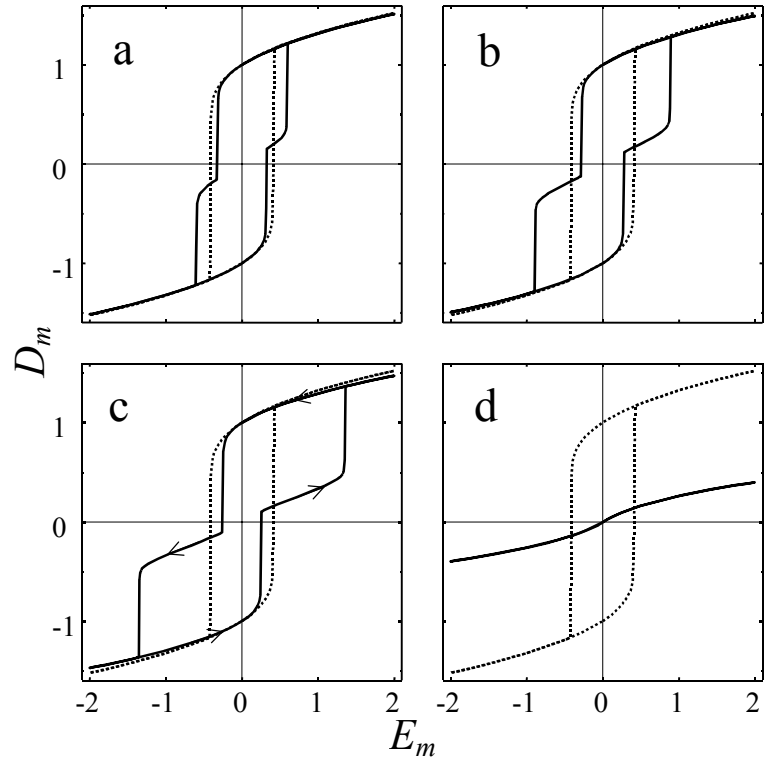

**Figure 5**. The dependence of dimensionless induction on dimensionless field (hysteresis loops) for frequency $w$=0.001, external field amplitude $E_{mA}$=2 and different values of $R^2$=0.1, 0.3, 0.5, 1 (parts a, b, c, d respectively). Initial conditions for the different solid curves are $\Delta^2_D(\tau=0)$=0, $K_{Dp}(\tau=0)$=0, $D_m(\tau=0)$=1 (parts a, b, c,) and $D_m(\tau=0)$=0 (part d). The dotted curve corresponds to the "classic" loop in the pure ferroelectric ($R$=0) at the same frequency value.

The hysteresis loops represented in Figs. 7 and 8 are obtained with the same parameters as in Fig. 5, but with the higher frequency values. It is seen that Fig. 7 is qualitatively similar to the Fig. 5: there are the constrictions on hysteresis loops for $R^2 \leq 0.5$, but for $R$=1 only minor loop exists and paraelectric-like dependence $D(E)$ (see Fig. 5d) cannot be achieved. With the frequency increase the constrictions are smeared and loops tend to the ones of pure ferroelectrics (see dotted lines in Fig. 8). Also under the frequency increase, minor loops arise for the smaller values of parameter $R$ at the same external field amplitude (compare Fig. 7c and Fig. 8c).

The decrease of the external electric field amplitude leads to the strong deformations of the hysteresis loop even for the pure ferroelectrics and its subsequent disappearance. Ellipse-like form of the curves can be explained by the dielectric losses. For the smaller external field amplitude only the loops of linear dielectrics with losses exist Thus for the ferroelectrics with impurity concentration fluctuations the disordered state is stable in the small enough external field.

Exactly the switching between ordered and disordered states in ferroelectrics with impurities causes both the minor hysteresis loops for the smaller external field and the constriction on loops for the larger external field amplitude. These phenomena can be easily explained on the basis of simple pictures of the free energy map evolution for different values of external field (see comments to the Fig. 2).

Therefore our theory predicts footprint and minor polarization hysteresis loops in ferroelectric materials with charged impurities and relatively high improper conductivity (e.g. the [1]-[3], [8], [10]). It should be noted, that the origin of the constricted or double loops in aged ferroelectric ceramics BaTiO₃ and (Pb,Ca)TiO₃ without charged impurities is caused by the mechanical clamping of spontaneous polarization switching [22] and so it lies outside of our theoretical consideration.

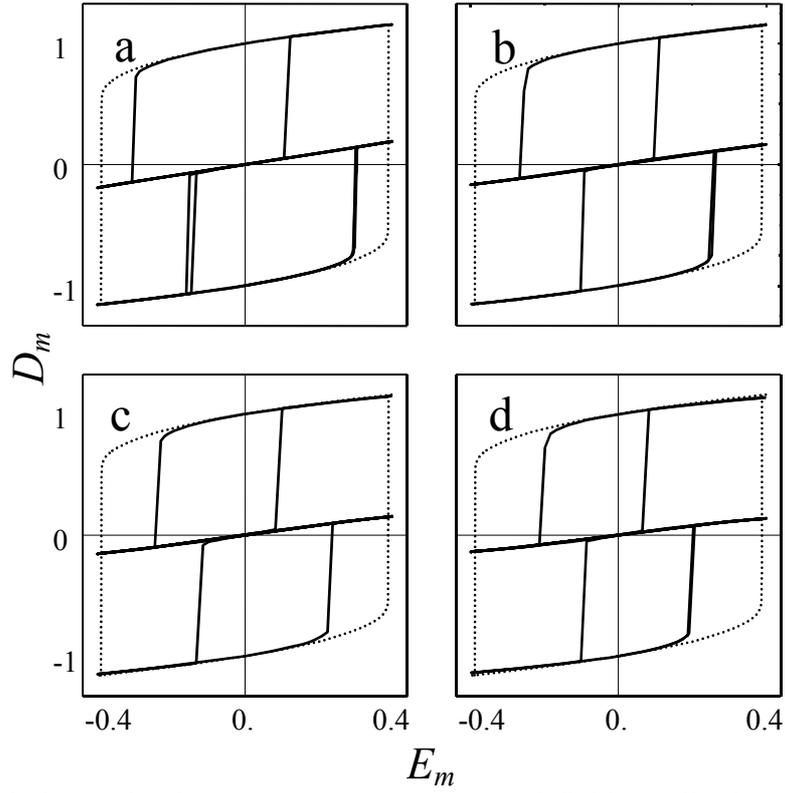

**Figure 6**. Hysteresis loops for frequency $w$=0.001, external field amplitude $E_{mA}$=0.4 and different values of $R^2$=0.1, 0.3, 0.5, 1 (parts a, b, c, d respectively). Initial conditions are $\Delta^2{}_D(\tau=0)$=0, $K_{Dp}(\tau=0)$=0, $D_m(\tau=0)$=-1. The dotted curve corresponds to the "classic" loop in the pure ferroelectric ($R$=0) at the same frequency value.

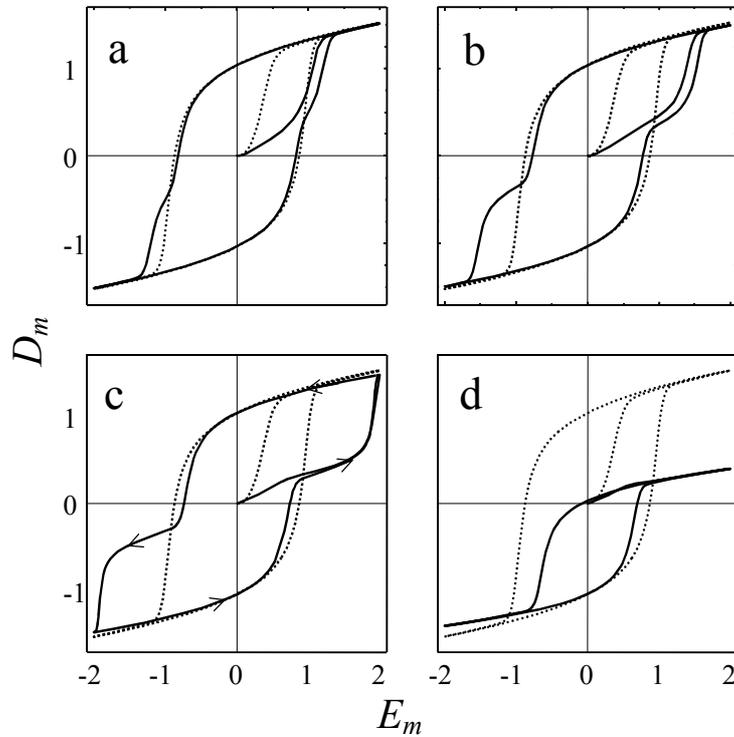

**Figure 7**. Hysteresis loops for frequency $w$=0.1, external field amplitude $E_{mA}$=2 and different values of $R^2$=0.1, 0.3, 0.5, 1 (parts a, b, c, d respectively). Initial conditions are $\Delta^2{}_D(\tau=0)$=1, $K_{Dp}(\tau=0)$=0, $D_m(\tau=0)$=0. The dotted curves corresponds to the "classic" loops in the pure ferroelectric ($R$=0) at the same frequency value.

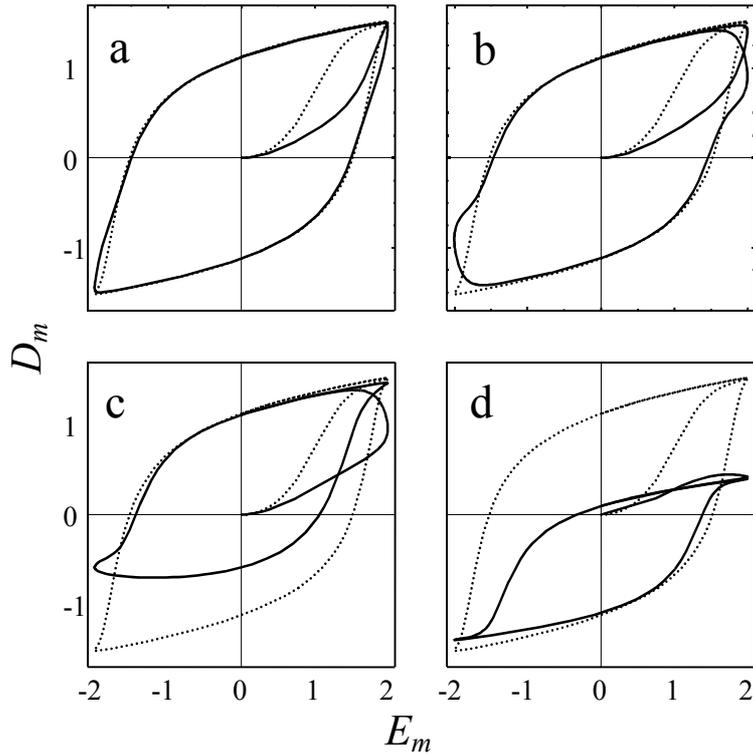

**Figure 8**. Hysteresis loops for frequency $w$=0.5, external field amplitude $E_{mA}$=2 and different values of $R^2$=0.1, 0.3, 0.5, 1 (parts a, b, c, d respectively). Initial conditions are $\Delta^2_D(\tau=0)$=1, $K_{Dp}(\tau=0)$=0, $D_m(\tau=0)$=0. The dotted curves correspond to the "classic" loop in the pure ferroelectric ($R$=0) at the same frequency value.

## DICUSSION

We have proposed the phenomenological description of polarization switching peculiarities in some ferroelectric semiconductor materials with charged defects. It is shown that the impurity concentration fluctuations lead to the ferroelectric disordering of the considered system. The quantitative degree of this disorder is the parameter $\sqrt{\overline{\delta D^2}}$ characterizing the inhomogeneity of the induction distribution. Mean induction $\overline{D}$ is the order parameter. For the first time the system of coupled equations (24) that determines the evolution of these parameters has been derived.

Solving the system of coupled equations one can get the information about system ordering as a whole, without defining concrete space distribution of the appeared inhomogeneities, domain walls characteristics, correlation radius of Cross regions or sizes of originated microdomains. In order to obtain this kind of information one has to solve the system of equations (13)-(16) with the specified distribution of impurity concentration fluctuations $\delta\rho_s(\mathbf{r})$, but the consideration of this problem was not the purpose of the present paper.

We would like to underline, that in contrast to the system (13)-(16) the averaged system of the coupled equations (24) does not contain any information about induction gradient across the sample. This happened rather due to the local compensation of the strong inhomogeneous electric field in the vicinity of charged defects by the movable charge carriers, then due to enough sample thickness in order to neglect the size effects and the depolarization field influence [23], [24]. Also we suppose that inhomogeneous mechanical stresses arisen near defects are small enough or compensated by the sample treatment.

According to our theory random inhomogeneities in the defect distribution throughout the sample leads to the stabilization of disordered state ($\sqrt{\overline{\delta D^2}} >> \overline{D}$). For the small amplitudes of external field this state reveals itself as switching from the ordered state to the disordered one (the so-called minor loop, see e.g. [1]). When the external field increases this minor loop transforms into the loop with constriction or footprint-type hysteresis loop [9]. In this case the constriction corresponds to the switching from ordered state to disordered one and then again to the ordered state with the opposite direction of the induction.

Ferroelectric hysteresis loops with the constrictions or footprint loops are observable in some ferroelectric materials. For example, footprint loops exist in the plumbum zirconate-titanate ceramics doped with La [10], namely in $Pb_{1-3y/2}La_yZr_xTi_{1-x}O_3$ at x=0.35, y=0.08, 0.084 and x=0.3, y=0.076, 0.079, which is regarded as relaxor material. In this material La ions have excess charge and can be regarded as charged defects. Notice, that our theory predicts transformation from footprint to the minor loop with the external field frequency increase as it was observed in SBN single crystals doped with cerium [2]. It is clear that our theory describes qualitatively minor loop but not the aging process seen as the loop degradation. This may be related to the fact that neither finite domain wall thickness, nor possible evolution of the charge fluctuations caused by the relaxation/origin of internal stresses around defects was taken into account in our simplified model. These problems as well as the calculation of the system dielectric response are in progress now.

We can conclude, that coupled equations (24) qualitatively describe the polarization switching and ferroelectric disordering caused by charged defects in such bulk ferroelectric materials with improper conductivity as relaxor PLZT-ceramics and SBN single crystals doped with donor impurities La and Ce correspondingly.

## Acknowledgments.

The authors are greatly indebted to Prof. N.V. Morozovsky for frutfull discussions of the model and useful remarks to the manuscript.

## APPENDIX A

Let us express the field variation $\delta E_z$ can be via $\delta D$ and $\delta \rho_s$. In accordance with (21b) and (20) one obtains that $\delta n \approx (1/4\pi)\partial \delta D/\partial z - \delta \rho_s$. Having substituted this expression into (21a), one obtains (23) with the help of inequality (22), namely:

$$\delta E_z \approx \left(E_0(t) - \frac{\kappa}{\mu}\frac{\partial}{\partial z}\right)\left(\frac{1}{4\pi\overline{\rho}_s}\frac{\partial}{\partial z}\delta D - \frac{\delta\rho_s}{\overline{\rho}_s}\right) \approx \frac{\kappa}{\mu}\frac{\partial}{\partial z}\left(\frac{\delta\rho_s}{\overline{\rho}_s}\right) - E_0(t)\frac{\delta\rho_s}{\overline{\rho}_s}. \quad (A.1)$$

The equations for $\overline{\delta D^2}$ and $\overline{\delta D\delta\rho_s}$ obtained directly from (16) have the form:

$$\frac{\Gamma}{2}\frac{\partial}{\partial t}\overline{\delta D^2} + \left(\alpha + 3\beta\overline{D}^2(t)\right)\overline{\delta D^2} + \beta\overline{\delta D^4} = \gamma\overline{\delta D\frac{\partial^2\delta D}{\partial \mathbf{r}^2}} + \overline{\delta D\delta E_z}, \quad (A.2)$$

$$\Gamma\frac{\partial}{\partial t}\overline{\delta D\delta\rho_s} + \left(\alpha + 3\beta\overline{D}^2(t)\right)\overline{\delta D\delta\rho_s} + \beta\overline{\delta D^3\delta\rho_s} = \gamma\overline{\delta\rho_s\frac{\partial^2\delta D}{\partial \mathbf{r}^2}} + \overline{\delta\rho_s\delta E_z}. \quad (A.3)$$

One can derive from (A.1) the following approximations for the correlations:

$$\overline{\delta D\delta E_z} = -E_0(t)\frac{\overline{(\delta\rho_s\delta D)}}{\overline{\rho}_s} + \frac{\kappa}{\mu}\overline{\left(\delta D\frac{\partial}{\partial z}\frac{\delta\rho_s}{\overline{\rho}_s}\right)} \approx -E_0(t)\frac{\overline{(\delta\rho_s\delta D)}}{\overline{\rho}_s}. \quad (A.4)$$

In (A.4) the term $\frac{\kappa}{\mu}\overline{\left(\delta D\frac{\partial}{\partial z}\frac{\delta\rho_s}{\overline{\rho}_s}\right)} \sim \frac{\kappa}{\mu\overline{\rho}_s d}\overline{(\delta\rho_s\delta D)} = \frac{4\pi\varepsilon_\perp R_D^2}{d}\overline{(\delta\rho_s\delta D)}$ can be neglected under the assumption that the screening of impurity ions is strong enough to satisfy the inequality $R_D^2/d^2 << E_0/\left(4\pi\varepsilon_d\overline{\rho}_s\right)$ (see (20)). For a thick sample with equivalent boundaries $z = \pm\ell$ we obtain from (A.1) that

$$\overline{\delta\rho_s\delta E_z} = -E_0(t)\frac{\overline{\delta\rho_s^2}}{\overline{\rho}_s} + \frac{\kappa}{2\mu}\overline{\left(\frac{\partial}{\partial z}\frac{\delta\rho_s^2}{\overline{\rho}_s}\right)} \equiv -E_0(t)\frac{\overline{\delta\rho_s^2}}{\overline{\rho}_s}. \quad (A.5)$$

Taking into account (20) one obtains that

$$\overline{\gamma \delta D \frac{\partial^2 \delta D}{\partial \mathbf{r}^2}} = -\gamma \overline{\left(\frac{\partial \ \delta D}{\partial \mathbf{r}}\right)^2} \cong -\frac{\gamma}{d^2} \overline{\delta D^2}, \qquad \left|\frac{\gamma}{d^2} \overline{\delta D^2}\right| << \left|\alpha\right| \overline{\delta D^2}, \qquad (A.6a)$$

$$\overline{\gamma \delta \rho_s \frac{\partial^2 \delta D}{\partial \mathbf{r}^2}} \sim -\frac{\gamma}{d^2} \overline{\delta \rho_s \delta D}, \qquad \left|\frac{\gamma}{d^2} \overline{\delta \rho_s \delta D}\right| << \left|\alpha \overline{\delta \rho_s \delta D}\right|, \qquad (A.6b)$$

and so gradient terms in (A.2)-(A.3) can be either neglected at $\gamma/d^2 << \alpha$ or the coefficient $\alpha$ can be renormalized as $\alpha \rightarrow \alpha_R = \left(\alpha + \gamma/d^2\right)$. Also one obtains from (9)-(10) that

$$\overline{\delta D^4} \approx \left(\overline{\delta D^2}\right)^2, \qquad \overline{\delta D^3 \delta \rho_s} \approx \overline{\delta D^2} \ \overline{\delta D \delta \rho_s}. \qquad (A.7)$$

Using (A.3)-(A.6) we obtain the equations (26b) and (26c) from the equations (A.1) and (A.2) if only $\gamma/d^2 << \alpha$ (see (20)).